\documentclass[10pt,letterpaper,twocolumn]{article} 

\usepackage{ol2}
\usepackage[draft]{hyperref}
\usepackage{amsmath}

\begin{document}

\twocolumn[ 

\title{Nematicons in liquid crystals with negative dielectric anisotropy}


\author{Jing Wang$^1$, Junzhu Chen$^1$, Jinlong Liu$^{1,2}$, Yiheng Li$^1$, Qi Guo$^{1,4}$, Wei Hu$^{1,5}$, Dayu Li$^3$, Yonggang Liu$^3$, and Li Xuan$^3$}

\address{
$^1$Laboratory of Nanophotonic Functional Materials and Devices, South China Normal University, Guangzhou 510006, China\\
$^2$College of Science, South China Agricultural University, Guangzhou, 510642, China\\
$^3$State Key Laboratory of Applied Optics, Changchun Institute of Optics, Fine Mechanics and Physics, Chinese Academy of Sciences, Changchun 130033, China\\
$^4$e-mail:guoq@scnu.edu.cn\\
$^5$e-mail:huwei@scnu.edu.cn
}

\begin{abstract}We report a theoretical and experimental work on the nematicon in the planar cell containing the nematic liquid crystal with negative
dielectric anisotropy, aligned homeotropically in the presence of an externally applied voltage. The formation of the soliton is resulted from the balance between the linear difrraction and the nonlocal nonlinearity due to molecular reorientation.
\end{abstract}


 ] 

\noindent

\bigskip

Nematicons, spatial optical solitons in nematic liquid crystals (NLC), have been the subject of intense theoretical and experimental studies over the past two decades \cite{Peccianti2012}. The pioneering work on nematicons was reported in 1993 by Braun {\it et al.} \cite{Braun1993}. They investigated the strong self-focusing of a laser beam in NLC in various geometries, from which they recognized the importance of molecular reorientation and anchoring at the boundaries. Subsequently, in 1998, Warenghem {\it et al.} observed the beam self-trapping in capillaries filled with dye-doped NLC \cite{Warenghem1998}. In the same year, Karpierz {\it et al.} observed the same phenomenon in planar cells with homeotropically aligned NLC \cite{Karpierz1998}. They lowered the required power of observing nematicons to milliwatt levels. In 2000, Peccianti {\it et al.} reported on nematicon formation in planar cells containing a NLC aligned homogeneously in the presence of an externally applied voltage \cite{Peccianti2000}. They extended the propagation length of nematicons for millimeter levels and found the adequate model for describing nematicon propagation.

The investigation on nematicon was sparse until Conti {\it et al.} found that the NLC with a pretilt angle induced by an external
low-frequency electric field is a kind of strongly nonlocal nonlinear medium and nematicons are a kind of accessible solitons \cite{Conti2003, Conti2004, Snyder1997}. They derived a simplified model and linked nematicons with quadratic solitons. The basic properties on nematicon have been revealed gradually ever since. Among others, we have to mention the interactions between two nematicons \cite{Peccianti2002, Hu2006, Hu2008, Skuse2008}. Recently, Piccardi {\it et al.} reported the dark nematicon formation in planar cells filled with dye-doped NLC aligned homeotropically, which can provide an effective negative nonlinearity \cite{Piccardi2011}. The negative nonlinearity is realized through the guest-host interaction.

In this letter, we observed the nematicon formation in planar cells containing a NLC  with negative dielectric anisotropy and positive optical anisotropy aligned homogeneously in the presence of an externally applied voltage. Following the method in \cite{Conti2003}, we got a simplified model with a negative Kerr coefficient and an oscillatory periodic response function, which can support bright nematicons. We outlined the connection between the simplified model and the equations describing quadratic solitons \cite{Buryak1995, Nikolov2003, Esbensen2012, Wang2014}.

We performed a series of experiments to observe the nematicon formation in NLC with negative dielectric anisotropy. The experimental setup is illustrated in Fig. 1(a). A light beam from a Verdi laser was focused by a 10X microscope objective and launched into a 80-$\mu m$-thick NLC cell. The configuration of the cell was shown in Fig. 1(b) and the cell was filled with the KY19-008 NLC, whose $n_\parallel=1.726$, $n_\perp=1.496$, average elastic constant $K=1\times10^{-11}N$, optical anisotropy $\epsilon^{op}_a=0.74106$, and dielectric anisotropy $\epsilon^{rf}_a=-5.3$. Owing to the negative dielectric anisotropy, the NLC molecules will try to adjust in a low-frequency applied electrical field in such a manner that the molecule axes turn perpendicular to the direction of the electric field \cite{Schiekel1971}. A microscope and a CCD camera were used to collect the light scattered above the cell during propagation.
\begin{figure}[tp]
\centerline{\includegraphics[width=8.4cm]{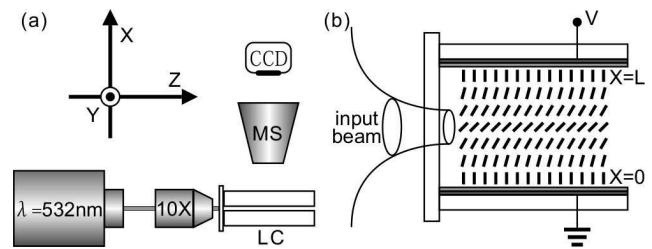}}
\caption{(Color online) Sketch of the experimental setup (a) and homeotropically aligned nematic liquid crystal cell (b) for the observation of nematicons.}
\end{figure}

One group of experimental results are shown in Figs. 2(a) and 2(c). The launched power and width for each beam is fixed to 4.42 mW and 4 $\mu m$ when the bias is changed. Fig. 2(a) shows the linear diffraction in absence of the bias and Fig. 2(c) shows the nemation formation at $V=3.4\text{V}$ above  the Fr\'{e}edericks threshold \cite{Peccianti2004, Ruan2008}
\begin{equation}\label{eq1}
V_{fr}=\pi(\frac{K}{\varepsilon_0|\varepsilon_a^{rf}|})^{1/2},
\end{equation}
where $\epsilon_0$ is the vacuum permittivity. Introducing the value of $\epsilon^{rf}_a$ and $K$ for KY19-008 NLC, we can get $V_{fr}\approx1.45\text{V}$.
The physical mechanism of the nonlinearity in NLC is optically induced molecular reorientation. If the bias is less than the Fr\'{e}edericks threshold, the NLC molecules will not turn. Then the optical beam will not reorientation the NLC molecules because of the optical electric field is weak under the power in our experiment. Therefore, the optical beam will diffract in absence of the bias and form soliton only if the bias is greater than the Fr\'{e}edericks threshold.
Figs. 2(b) and 2(d) show the numerical results for contrast, which are calculated based on Eqs. (\ref{eq2})--(\ref{eq3}) with Gaussian beam as an incident profile.
\begin{figure}[tp]
\centerline{\includegraphics[width=8.4cm]{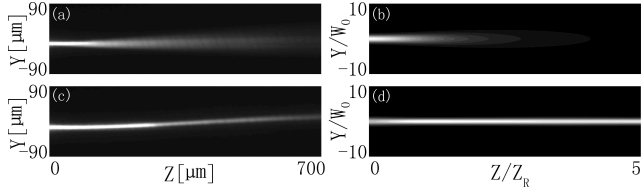}}
\caption{(Color online) $X$-polarized $e$-beam propagation with power $P=4.42mW$ and width $w=4\mu m$ for different voltage. (a) and (b) Diffraction in the absence of voltage bias. (c) and (d) Nematicon formation in the presence of 3.4 V and 3.01V bias at 1 kHz, respectively.}
\end{figure}

In the presence of an externally applied (low-frequency) electric field $E_{rf}$,  the evolution of the slowly varying envelope $A$ of a paraxial optical beam linearly polarized along $X$(an extraordinary light) and propagating along $Z$ can be described by the system\\
\begin{equation}\label{eq2}
2ik\frac{\partial A}{\partial Z} + \nabla^2_{XY} A+k^2_0 \epsilon^{op}_a (\sin^2\theta-\sin^2\theta_0) A=0,
\end{equation}

\begin{equation}\label{eq3}
2K(\frac{\partial^2\theta}{\partial Z^2}+\nabla^2_{XY} \theta) +\epsilon_0(\epsilon^{rf}_a E^2_{rf}+\epsilon^{op}_a\frac{|A|^2}{2})\sin(2\theta) =0,
\end{equation}\\
where $\theta$ is the tilt angle of the NLC molecules, $\theta_0$ is the nadir tilt in the absence of light, $k = k_0n_e(\theta_0)$ with $k_0$ the vacuum wavenumber and $n_e(\theta_0)=n_\perp n_\parallel/(n_\parallel^2cos^2\theta_0+n_\perp^2sin^2\theta_0)^{1/2}\approx(n_\perp^2+\epsilon^{op}_asin^2\theta_0)^{1/2}$ the refractive index of the extraordinary light at $\theta_0$, $\nabla^2_{XY}=\partial_X^2+\partial_Y^2$, $\epsilon^{rf}_a=\epsilon_\parallel-\epsilon_\perp(<0)$, $\epsilon^{op}_a=n_\parallel^2-n_\perp^2(>0)$. The term $\partial^2_Z\theta$ in Eq.(\ref{eq3}) was proven to be negligible compared to $\nabla^2_{XY} \theta$, therefore it can be removed. The homeotropical boundaries and anchoring at the interfaces define $\theta|_{X=0}=\theta|_{X=L}=\pi/2$, where $L$ is the cell thickness. In the absence of light, the pretilt angle $\hat{\theta}$ is symmetric along $X$ about $X = L/2$ (the cell center) and depends only on $X$: \\
\begin{equation}\label{eq4}
2K\frac{\partial^2\hat{\theta}}{\partial X^2} +\epsilon_0\epsilon^{rf}_a E^2_{rf}\sin(2\hat{\theta}) =0.
\end{equation} \\

Furthermore, we can set $\theta=\hat{\theta}+(\hat{\theta}/\theta_0)\Phi$, with $\Phi$ being the optically induced perturbation. Noting that $\hat{\theta}\approx\theta_0$ and $\partial_X\hat{\theta}\approx0$ in the middle of the cell when the beam width is far smaller than the cell thickness, we can simplify Eq.(\ref{eq2}) and Eq.(\ref{eq3}) into the following system, which describes the coupling between $A$ and $\Phi$:\\
\begin{equation}\label{eq5}
2ik\frac{\partial A}{\partial Z} + \nabla^2_{XY} A+k^2_0 \epsilon^{op}_a \sin(2\theta_0) \Phi A=0,
\end{equation}

\begin{equation}\label{eq6}
w_m^2\nabla^2_{XY} \Phi+\Phi -\frac{2n_0n_2}{\epsilon^{op}_a\sin(2\theta_0)}|A|^2=0,
\end{equation}\\
where the parameter $w_m(w_m>0$ for $|\theta_0| \leq \pi/2)$, that is, the characteristic length of the nonlinear response function, reads:\\
\begin{equation}\label{eq7}
w_m=\frac{1}{E_{rf}}\{\frac{2\theta_0K}{\epsilon_0|\epsilon^{rf}_a|\sin(2\theta_0)[1-2\theta_0\cot(2\theta_0)]}\}^{1/2},
\end{equation}\\
and \begin{equation}\label{eq8}
n_2=-\frac{(\epsilon^{op}_a)^2\theta_0\sin(2\theta_0)}{4n_0|\epsilon^{rf}_a|E_{rf}^2[1-2\theta_0\cot(2\theta_0)]}.
\end{equation}\\
The nonlinear refractive index coefficient $n_2$ is defined as suggested by Peccianti et al.\cite{Peccianti2004}. It must be noted, however, that $n_2$ is negative for the liquid crystal with negative dielectric anisotropy,
while it is positive for that with positive dielectric anisotropy.
\\
\\
\begin{figure}[tp]
\centerline{\includegraphics[width=8.4cm]{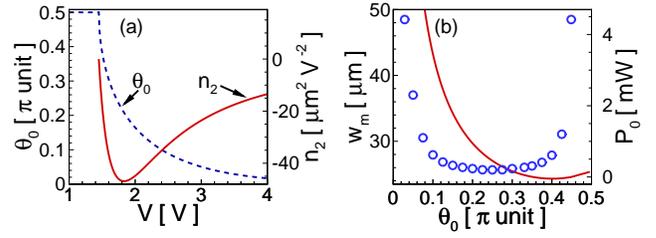}}
\caption{(Color online) (a) The pretilt angle $\theta_0$ and the Kerr coefficient $n_2$ of the NLC vs the bias voltage $V$. (b) The characteristic length $w_m$ (a solid curve) and the critical power of a single soliton (circles) vs the pretilt angle $\theta_0$. The parameters are for a $80-\mu m$-thick cell filled with the NLC (KY19-008) and the critical power is a numerical result.}
\end{figure}

A monotonous function of $\theta_0$ on $E_{rf}$ (or $V$ by $V=E_{rf}L$) is described by Eq. (\ref{eq4}). As shown in Fig. 3(a), $\theta_0$ decreases monotonously from $\pi/2$ to $0$ with increasing the bias above the Fr\'{e}edericks threshold. For $E_{rf}$ higher than the Fr\'{e}edericksz threshold, the approximation
\begin{equation}\label{eq9}
\theta_0\approx\frac{\pi}{2}(\frac{E_{fr}}{E_{rf}})^3
\end{equation}
is satisfactory, where $E_{fr}=V_{fr}/L$. Therefore, we can clearly see from Eqs.(\ref{eq7}) and (\ref{eq8}) that $w_m$ and $n_2$ are determined by $V$ or $\theta_0$ for a given NLC cell configuration. As shown in Fig. 2, $w_m$ and $n_2$ changes nonmonotonously with increasing $\theta_0$ and $V$, respectively. Both of them have a minimum value. There we also show the relation of the critical power $P_0$ on $\theta_0$ calculating numerically based on Eqs. (\ref{eq2})--(\ref{eq3}) with Gaussian beam as an incident profile.

Introducing the normalization that $x=X/w_m$, $y=Y/w_m$, $z=Z/(kw_m^2)$, $u=A/A_0$, $\phi=\Phi/\Phi_0$, where $A_0=[8K/k_0^2w_m^4\epsilon_0\epsilon^{op2}_a\sin^2(2\theta_0)]^{1/2}$, $\Phi_0=2/[k_0^2w_m^2\epsilon^{op}_a\sin(2\theta_0)]$, we have the dimensionless system,
\begin{equation}\label{eq10}
i\frac{\partial u}{\partial z} + \frac{1}{2}\nabla^2_{\perp} u+\phi u=0,
\end{equation}

\begin{equation}\label{eq11}
\nabla^2_{\perp} \phi + \phi = -|u|^2,
\end{equation}
where $\nabla^2_{\perp}=\partial_x^2+\partial_y^2$. We consider a planar geometry with the boundary condition $\phi|_{x=0, l}=0$. Eq.(\ref{eq11}) has a particular solution in the form of a convolution integral of $|u|^2$ with the function $R$:\\
\begin{equation}\label{eq12}
\phi(x,y)=\int_{0}^{l}\int_{-\infty}^{+\infty}R(x,y;x^\prime,y^\prime)|u(x^\prime,y^\prime)|^2dx^\prime dy^\prime.
\end{equation}
Here
\begin{equation}\label{eq13}
R(x,y)= \sum_{m=1}^{\infty}a_m(y)\sin\frac{m \pi x}{l},
\end{equation}
with
\begin{equation}\label{eq14}
a_m= \begin{cases} -\frac{\sqrt{\varsigma_m}}{l}\sin\frac{m\pi x^\prime}{l}\sin|\frac{y-y^\prime}{\sqrt{\varsigma_m}}| & m<l/\pi, \\
\frac{\sqrt{|\varsigma_m|}}{l}\sin\frac{m\pi x^\prime}{l}\exp[-|\frac{y-y^\prime}{\sqrt{|\varsigma_m|}}|] & m>l/\pi,
\end{cases}
\end{equation}
where $\varsigma_m=1/[1-(\frac{m\pi }{l})^2]$. This means that the tilt angle will be sine-oscillatory for a gaussian input beam when the planar thickness $l>\pi$ and exponential-decay $l<\pi$. From Eqs.(\ref{eq1}), (\ref{eq7}), and (\ref{eq9}), we can get
\begin{equation}\label{eq15}
l=\pi^{4/3}(2\theta_0)^{-5/6}\sin^{1/2}(2\theta_0)[1-2\theta_0\cot(2\theta_0)]^{1/2}.
\end{equation}
We can see that $l$ is changed only with $\theta_0$. The maximum of $l$ is comparable with $\pi$. Therefore, the tilt angle of original model Eqs. (\ref{eq2})--(\ref{eq3}) is not sine-oscillatory when $l>\pi$, although the tilt angle of simplified model Eqs. (\ref{eq10})--(\ref{eq11}) is.
\begin{figure}[tp]
\centerline{\includegraphics[width=8.4cm]{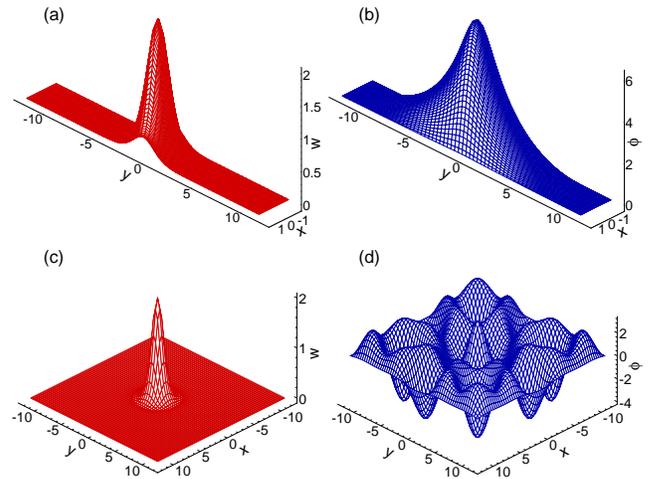}}
\caption{(Color online) The soliton profile [(a) and (c)] and tilt angle [(b) and (d)] for different sample size $l_x$ and $l_y$ and propagation constant $\beta$. (a)--(b)$l_x=3$, $l_y=23.6$, $\beta=4.3099$. (c)--(d)$l_x=l_y=23.6$, $\beta=1.4326$.}
\end{figure}

We found many solitons from Eqs. (\ref{eq10})--(\ref{eq11}) using numerical iteration method. It was found that the soliton profile is changed with the sample size $l_x$ and $l_y$ and the propagation constant $\beta$. Figs. 4 show the two group of typical soliton profile $w$ and the corresponding tilt angle $\phi$.
When $l_x<\pi$ (or $l_y<\pi$), $\phi$ is exponential-decay, as shown in Fig. 4(b). When $l_x>\pi$ (or $l_y>\pi$), $\phi$ is sine-oscillatory , as shown in Fig. 4(d). This result agrees with the above analysis of Greens function method.

In summary, we observed experimentally the nematicon formation in the planar cell containing the nematic liquid crystal with negative
dielectric anisotropy, aligned homeotropically in the presence of an externally applied voltage. We gave a theoretical model describing nematicon propagation, based on which we investigated the evolution of a Gaussian beam. And then we derived a simplified model, from which we found many bright nematicon with a sine-oscillatory response function and a negative Kerr coefficient.

This research was supported by the National Natural Science Foundation of China (Grant Nos. 11174090, 11174091, 11074080, and 11204299) and the Scientific Research Foundation of Graduate School of South China Normal University (Grant No. 2013kyjj014).

\end{document}